          [2001/04/25 1.1 (PWD)]
\documentclass[a4paper,twocolumn]{esapub} 
\usepackage{graphicx,natbib}

\title{HERSCHEL and Galaxies/AGN}
\author{Peter Barthel}
\affil{Kapteyn Institute, Univ. of Groningen,}
\affil{ P.O.~Box~800, NL--9700AV Groningen, The Netherlands}

\def\lea{\mathrel{\raise .4ex\hbox{\rlap{$<$}\lower 1.2ex\hbox{$\sim$}}}}
\def\gea{\mathrel{\raise .4ex\hbox{\rlap{$>$}\lower 1.2ex\hbox{$\sim$}}}}
\def \sol{\ifmmode _{\mathord\odot} \else $_{\mathord\odot}$ \fi}

\begin{document}


\maketitle

\begin{abstract}

Herschel will represent a breakthrough in the study of nearby gas-rich
and gas-poor galaxies, as it will for the first time permit imaging
photometric and spectroscopic observations of their ISM in the
FIR-submm wavelength range.  The unprecedented sensitivity and angular
resolution of Herschel will furthermore yield a breakthrough in our
understanding of distant galaxies and AGN, as their gas and dust --
both the ISM- and the AGN-related -- will for the first time come
within reach.  Herschel will undoubtedly yield major discoveries
concerning the cosmologically evolving gas and dust properties in
galaxies, back to very early epochs.

\end{abstract}

\section{Introduction}

Evidence is accumulating that star-formation in galaxies and the
occurrence of accretion-driven activity in their nuclei are
astrophysically connected processes. I'll give three prime
lines of reasoning.

1. The masses of the spheroidal components in galaxies grow together
with their massive black holes (MBHs) -- this is the well-known bulge
-- black hole correlation.

2.  Nuclear accretion is often preceded by a phase of enhanced
star-formation.  We suspected this from age dating of the host
galaxies of nearby Seyfert galaxies and QSO's (e.g.,
Gonz\'alez-Delgado et al. 2001, Canalizo \& Stockton 2001), but a
wonderful proof of the symbiosis has recently been obtained by
Kauffmann et al.  (2003), using SDSS data.  A remarkable plot from
this work is reproduced in Fig.~1: this emission line ratio diagram
shows the normal star-forming galaxy population, from low mass (left)
to high mass (bottom), with the "abnormal" populations of LINER and
Seyfert galaxies branching off towards the top right.  The Seyfert
galaxies, having high [O\,III] luminosities, were often found to display
young stellar populations, i.e., they are often post-starburst
systems.  {\it This is the plot all extragalactic astronomers -- the
privileged Herschel users in particular -- should be thinking about,
concerning its origin and its cosmic evolution!}

\begin{figure}
\centering
\includegraphics[width=0.8\linewidth]{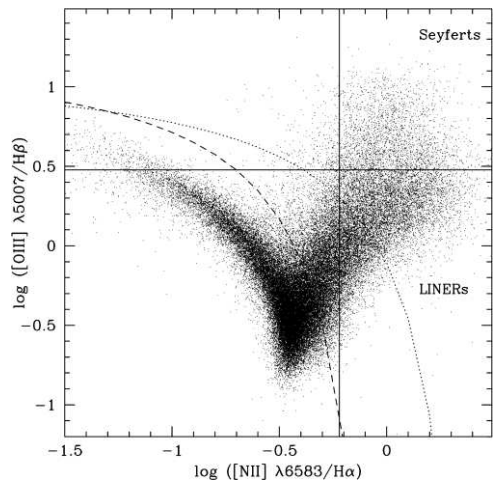}
\caption{Line-ratio plot for fifty-five thousand emission line 
galaxies, from the SDSS data base, reproduced with permission from 
Kauffmann et al. (2003). The dashed line separates the normal 
star-forming galaxies from the LINERs and the Seyferts.}
\end{figure}

3.  The resemblance of the cosmic star-formation history plot and the
QSO space density plot, in particular the $z\sim2$ peak, may not be a
coincidence.  Also the nature of the distant submm galaxies, most
likely containing a substantial AGN-fraction (e.g., Frayer et
al. 2004), supports an astrophysical link.

The following issues are also very relevant in this respect.

1. Given the fact that $z>6$ QSOs are being found, and that they possess
very massive black holes, the bulge -- black hole correlation just
mentioned implies that massive bulges {\it must} be in place, within a
Gigayear after the Big Bang.

2.  The emission line spectra of these $z>6$ QSOs differ little or
nothing from the spectra of more recent QSOs: the abundances in the
broad line gas are solar (or even super-solar).  This is indicating
that {\it processed} material is being accreted, and not pristine ISM
gas.

3. As reported in several recent studies, molecular gas is sometimes
found in distant AGN, both QSOs and radio galaxies, in staggering
quantities: there is plenty of fuel for star-formation.

In summary, it was long known that the two phenomena can co-exist
(e.g., NGC\,1068!), but the study of this important star-formation--AGN
connection evolving throughout cosmic time will be a major challenge
for extragalactic astronomers in the coming decade.  As will be
illustrated below, Herschel (and ALMA) can and will play an important
role in these investigations. Obviously this review cannot be complete
but it will hopefully raise interest from outside the already
impressively large Herschel communities.

\section{The legacy of IRAS and ISO (and Spitzer)}

As amply documented (e.g., Soifer et al. 1987), the all-sky IRAS
mission opened the dusty universe, detecting dust radiation from many
classes of extragalactic objects -- both expected and unexpected --
and classifying these dust properties. UltraLuminous InfraRed
Galaxies, ULIRGs, emerged as an important class of obscured, gas-rich
galaxies having their SEDs dominated by unusually strong far-infrared
emission.

Following in its footsteps, the observatory mission ISO subsequently
carried out more detailed investigations. ISO revealed the presence of
cold dust ($\sim$15\,K) in galaxies, and established the unexpected
fast decay of starbursts. Mid- and far-infrared luminosities, colors,
and temperatures were obtained for many objects in the local and
nearby ($z \lea 0.1$) universe. In nearby galaxies, ISO provided more
insight in the physics of the star-forming regions, in particular the
[C\,II] and [O\,I] cooling.  ISO imaging of the warm dust in nearby
galaxies proved to be a fine tool to address localized radiation
properties.  ISO revealed fascinating cases of obscured star-formation
such as in the Antennae NGC\,4038/4039.  Moreover, the spectroscopic
capabilities of ISO permitted great advance in our understanding of
the gas properties of nearby dusty and gas-rich objects. The
mid-infrared aromatic features were discovered, but also the
absorption characteristics, as well as the diagnostic emission lines
and the cooling lines in the mid- and far-IR. An excellent account of
the ISO legacy can be found in Genzel \& Cesarsky (2000).  NASA's
Spitzer mission is currently continuing this work, at much improved
sensitivity; impressive first results have appeared recently (ApJS
Vol. 154 No.1 issue), and some highlights were reported at this
meeting. Spitzer will surely bring great advancement in our
understanding of the spectral energy distributions (SEDs) of nearby
and distant galaxies, including the magnificent 2-D information
resulting from its infrared camera.

Equally interesting results are to be awaited from ASTRO-F, the
Japanese mission to be launched hopefully in 2005. ASTRO-F will
operate a sensitive large field-of-view near- and mid-IR camera,
and will in addition carry out an improved all-sky far-infrared
survey, FIS, with a predicted yield of over $10^7$ extragalactic
far-infrared sources -- a significant improvement with respect 
to IRAS. The ASTRO-F FIS data will undoubtedly be important for
Herschel.

\section{HERSCHEL EXPECTATIONS}

Broadly speaking, Herschel will continue where its predecessors
stopped.  However, in addition to the fascinating achievements which
one can foresee, Herschel will obviously make completely unexpected
discoveries: the telescope will point at known (active) objects but in
addition of course follow-up on its own discoveries.  Herschel will
extend the IRAS/ISO/Spitzer/ASTRO-F wavelength range into the submm,
at much greater sensitivity, at considerably higher spatial resolution
and at much higher spectral resolution.  These qualifications will
permit studying the mid- and far-infrared out to high redshifts. Given
that at these wavelengths star-formation and nuclear activity -- be
it indirectly -- are optimally revealed, Herschel will truly live up
to its Cornerstone expectations. The paragraphs below summarize the
key themes in this respect: dust and star-formation. {\it In short,
besides being the optimal far-IR spectrograph, Herschel will become
the perfect SED-machine.}

\subsection{DUST: nearby and far}

The optical--infrared SEDs for early and late type galaxies differ
most significantly in the mid- and far-IR. Whereas the former are
roughly flat (in the F$_{\nu}$ representation), or show a modest peak
around 2$\mu$m from stellar photospheric emission, the latter show a
peak around 100$\mu$m, indicative of cool dust. Starburst galaxies
such as M\,82 show that peak more pronounced and at somewhat shorter
wavelength ($\sim$~60 -- 80$\mu$m), while the SEDs of ULIRGs -- the
champion FIR-emitters -- are completely dominated by such a warm
far-infrared peak, the latter contributing around 90\% of the
bolometric emission. The strength of this warm dust component is
correlated with the optically thin synchrotron radio emission -- this
is the well known radio-FIR correlation (e.g., Condon 1992).

\begin{figure*}
\centering
\includegraphics[width=0.8\linewidth]{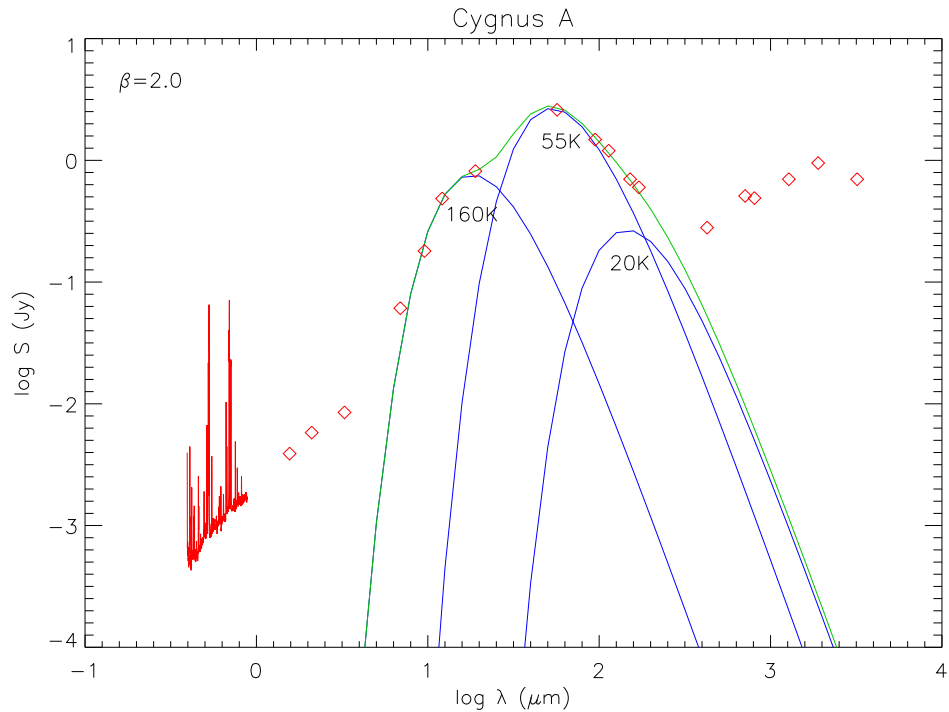}
\caption{The optical--radio SED for the Cygnus\,A radio galaxy 
host, with a three-component thermal dust decomposition (I. van Bemmel, 
priv. comm.)}
\end{figure*}

Far-infrared dust emission is also measured in active galaxies and
AGN.  The excess thermal emission in Seyfert-2 galaxy NGC\,1068 is
known since the 1970's, but the systematic properties of this emission
in Seyfert galaxies had to await the IRAS survey. Using IRAS data, De
Grijp et al.  (1985) were able to show that the infrared color
$\alpha^{60\mu}_{25\mu}$ provides a good AGN selector, that is to say:
besides star-formation related warm dust, active galaxies are
characterized with additional hot dust, which together yield
relatively flatter infrared spectral indices. ISO data permitted more
elaborate multi-component dust models, both for star-forming objects
(e.g., Klaas et al. 1997) and active galaxies (e.g., P\'erez
Garc\'{\i}a et al.  (1998), Spinoglio et al. 2002).  Moreover, the ISO
spectrographs allowed us to examine the richness of the mid- and
far-infrared spectra. These studies demonstrated the importance of the
mid- and far-infrared wavelength range, assessing the relative and
absolute strength of star-formation and nuclear activity, in a crude
way, for nearby objects.

Similar SED analysis tools were applied in case of more distant, more
powerful AGN, such as QSOs, radio galaxies, and radio-loud
quasars. These studies were only moderately successful due to
sensitivity constraints. Nevertheless, intriguing trends were reported
such as the dusty nature of QSOs (e.g., Haas et al. 2003) and the
star-formation connection (Van Bemmel \& Barthel 2001). ISO photometry
was also applied tot test AGN unification schemes (Van Bem\-mel et
al. 2000, Haas et al. 2004). Figure~2 illustrates the technique of
multi-component dust model fitting, in the (high S/N) case of the
nucleus of the Cygnus\,A radio galaxy.  Recalling the substantial
diffuse cold dust in galaxies, at least three components must be
present in active galaxies. The hot circumnuclear torus dust however
will display complicated aspect dependent optical thickness
effects. The latest torus models (Van Bemmel \& Dullemond 2003,
Nenkova et al. 2002) cannot yet be properly tested with current SED
data: high S/N SEDs for many classes of objects are eagerly awaited.

As mentioned already, the warm dust in galaxies draws from
star-formation (and also from torus emission). A strong warm dust
component can be inferred from infrared color-color plots and implies
relatively strong star-formation activity. This is ob\-viously true for
ULIRGs, prime cases of starburst activity, but also QSOs (with an
additional dust heating source, namely the AGN) can be put on such
plots.  Figure~3 indicates that indeed such star-forming AGN can be
identified: age-dating using optical spectra of the circumnuclear
populations in these QSOs (Canalizo \& Stockton 2001) supports their
post-starburst nature, cf. expectation. During the last few years
evidence is growing that QSO host galaxies may be richer in
star-formation fuel than initially thought (e.g., Evans et al. 2001,
Scoville et al. 2003).

Herschel will address the symbiosis\footnote{It is appropriate here to
recall a result from SED modeling studies by M.~Rowan-Robinson (1995):
``a QSO or Seyfert event appears to be inevitably accompanied by a
starburst''} of MBH accretion activity and star-formation out to high
$z$, using the PACS and SPIRE instruments.  This symbiosis is
manifested in the global FIR-submm spectral energy distributions as
well as in the detailed FIR-submm spectral properties.  Given its (one
to two orders of magnitude) improved sensitivity w.r.t. previous
missions, Herschel will be able to measure the SEDs and the
spectroscopic properties out to extreme redshifts, for many classes of
objects.  The dust content and composition of many types of distant
galaxies -- with and without AGN -- come within reach.  Analysis of
the SEDs must obviously take the latest generation multi-component
dust models into account.  Key issues include: dust in early type
galaxies, intra- and intercluster dust, dust in QSO hosts (are they
really mature quiescent ellipticals?), dust in radio galaxies, the
cosmologically evolving dust properties and dust-to-gas ratio in
galaxies, AGN unification aspects (including the search for obscured
type-2 AGN), the nature of sub-mm galaxies (see below), etcetera.  The
radio-FIR correlation will be an important tool in these
investigations, as it can help to isolate the star-formation related
dust component. Deep radio observations at long wavelengths with
sufficient angular resolution must provide the essential complimentary
data.

The spectroscopic capabilities of Herschel will be used to full
advantage for distant sources to obtain the near- and mid-IR
diagnostic line ratios which proved crucial to address the nature of
the energy source in nearby starburst and active galaxies (cf. Genzel
et al. 1998). The luminosities of critical low- and high ionization lines
but also PAH luminosities at intermediate and high redshift can be obtained
from $R\sim100 - 200$ spectroscopy in about ten hours of exposure time.

\begin{figure}
\centering
\includegraphics[width=0.8\linewidth]{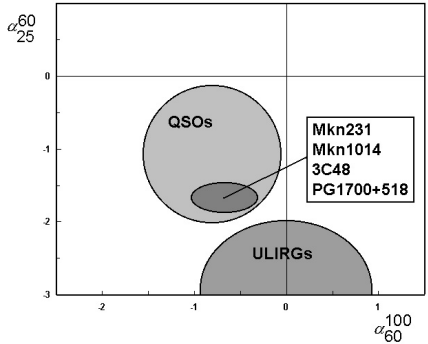}
\caption{Color-color diagram for QSOs and ULIRGs, with the location of 
four gas-rich, post-starburst QSO's, including the radio-loud 3C\,48 
indicated.Indeed the post-starburst QSOs lie close to the ULIRGs.}
\end{figure}

Submm galaxies may well be the distant ULIRGs and their contribution
to the infrared-submm background is significant (e.g., Blain et
al. 2002). As such, examination of their SEDs and assessment of their
energetics is a prime task, which Spitzer has begun, but which
Herschel will do better, given its larger wavelength coverage. Herschel
photometry in the far-infrared can be combined with ALMA mm-photometry
to permit dust photo-$z$ assessment (Yun \& Carilli 2002). Herschel
may well turn out to be the only instrument capable of determining the
precise redshift of obscured submm galaxies, from their molecular or
ionic cooling lines .... This will require tens of hours of integration
but is not unfeasible.

\subsection{STAR-FORMATION: present and past}

ISO spectra have brought great advancement in our understanding of the
star-formation process, both locally in the Milky Way and globally in
nearby and low redshift galaxies. Spectra taken with the Spitzer IRS
are currently refining this knowledge and extending the distance
coverage. The improved angular resolution of Herschel nevertheless is
eagerly awaited for investigations of the details of the
star-formation process: PDR cooling, the mass function of Giant
Molecular Clouds, the fragmentation, properties of Super Star
Clusters, etc. Herschel's HIFI instrument will explore the chemistry
of starburst evolution and differentiation, in the Galaxy and in
nearby galaxies. Template cooling line spectra containing the many CO,
H$_2$O, and related lines will be obtained, and their application for
distant star-forming objects will be investigated. The circumnuclear
X-ray dominated regions (XDRs) in nearby active galaxies will be 
investigated, again yielding template spectra. PACS and SPIRE will
make imaging studies of the molecular cloud complexes, both in
continuum and line (including PAH) radiation.  PACS will obtain
imaging spectroscopy of star-forming regions, focussing on the
detailed physics of Galactic PDRs and the star-formation--AGN
interplay in nearby (active) galaxies.

Concerning cooling, a lot of work must be done. Cooling through
[C\,II]$\lambda158\mu$ and [O\,I]$\lambda63\mu$ is not well
understood, and the [C\,II] deficit in strongly star-forming galaxies
and ULIRGs (e.g., Malhotra et al. 1997) has now been measured at extreme
redshift (Bolatto et al. 2004).

Taking full advantage of the Herschel sensitivity, star-formation in 
early type galaxies will be investigated. Also the puzzling class of 
blue compact dwarfs will be studied: what is the origin of the
low metallicity of these objects?

Herschel spectroscopic imaging observations may shed light on the
driving force of nuclear activity: can we finally solve the bar issue,
is there a direct physical link with circumnuclear star-formation?
Also close to home, the details of the radio-FIR correlation,
including the small-scale break-down can be studied using Herschel
imaging. Such studies will refine the already impressive Spitzer
studies (e.g., Gordon et al. 2004).

Concerning distant and powerful starburst galaxies such as HyLIRGs,
submm galaxies, Lyman-break galaxies, and high $z$ ellipticals,
Herschel will undoubtedly quantify the level of star-formation and
settle the question as to the presence of additional energy
sources. With help from the negative K-correction, galaxies having
star-formation rate $\sim~10^2$ M$_{\sol}$/yr can be observed out to
$z \sim 2$, whereas SFRs of order $10^3$ M$_{\sol}$/yr can be traced
out to much higher redshifts. Large Herschel Key Programs will be
devoted to a census of the evolving starburst galaxy population, on
the basis of data from the the large-area Herschel photometric
surveys. Indeed, these surveys will address the question ``how do
galaxies form?'' Together with ALMA and other ground-based mm
telescopes Herschel will determine the cosmologically evolving
star-formation efficiency SFE = L$_{\rm FIR}$/L$_{\rm gas}$.
Star-formation in very high redshift AGN, such as radio galaxies
(e.g., Greve et al. 2004) and QSOs (e.g, Walter et al. 2003) will
undoubtedly receive much Herschel attention.

\section{Summary}

Star-formation and AGN activity were in the past at least one order of
magnitude more prominent than they are now. Understanding these
evolving phenomena and their interplay is a key theme of today's
astrophysics. To obtain that understanding is the raison d'\^etre of
the Herschel Space Observarory.  Herschel will deal with star and
galaxy formation, throughout cosmic history, globally and in detail:
MW, PDRs, GMCs, SSCs, BCDs, Es, LBGs, EROs, DRGs, SMGs, XDRs, QSOs,
PRGs, LIRGs, ULIRGs and HyLIRGs, .... \footnote{Milky Way,
PhotoDissociation Regions, Giant Molecular Clouds, Super Star
Clusters, Blue Compact Dwarfs, Ellipticals, Lyman-Break Galaxies,
Extremely Red Objects, Distant Red Galaxies, Sub-Millimeter Galaxies,
X-ray Dominated Regions, Quasi-Stellar Objects, Powerful Radio
Galaxies, Luminous InfraRed Galaxies, UltraLuminous InfraRed Galaxies,
HYperLuminous InfraRed Galaxies, .....} you name it! These are
exciting times and Herschel is going to make it even more
exciting. Those who wish to join in the planning of Herschel studies
are referred to the Herschel website {\tt
http://www.rssd.esa.int/Herschel/}.

\section*{Acknowledgements}

The author acknowledges input from the Herschel Instrument Science
Teams.  Thanks also to Guinevere Kauffmann and Ilse van Bemmel,
for permission to reproduce Figs.~1 and 2, and to Marco Spaans for 
comments on the manuscript.

\section*{References}

Blain A.W., Smail I., Ivison R.J., et al., 2002, Phys. Rep. 369, 111 

Bolatto A.D., di Francesco J., Willott C.J., 2004, ApJ 606, L101

Canalizo G., Stockton A., 2001, ApJ 555, 719

Condon J.J., 1992, ARA\&A 30, 575

De Grijp M.H.K., Miley G.K., Lub J., et al., 1985, Nature 314, 240

Evans A.S., Frayer D.T., Surace J.A., et al., 2001, AJ 121, 3285

Frayer D.T., Chapman S.C., Yan L., et al., 2004, ApJS 154, 137

Genzel R., Lutz D., Sturm E., et al., 1998, ApJ 498, 579

Genzel R., Cesarsky C.J., 2000, ARAA 38, 761

Gonz\'alez Delgado R.M., Heckman T., Leitherer C., 2001, ApJ 546, 845

Gordon K.D., P\'erez-Gonz\'alez P.G., Misselt K.A., et al., 2004, ApJS 154, 215

Greve T.R., Ivison R.J., Papadopoulos P.P., 2004, A\&A 419, 99

Haas M., Klaas U., M\"uller S.A.H., et al., 2003, A\&A 402, 87

Haas M., M\"uller S.A.H., Bertoldi F., et al., 2004, A\&A 424, 531

Kauffmann G., Heckman T.M., Tremonti C., et al., 2003, MNRAS 346, 1055

Klaas U., Haas M., Heinrichsen I., et al., 1997, A\&A 325, L21

Malhotra S., Helou G., Stacey G., et al., 1997, ApJ 491, L27

Nenkova M., Ivezi\'c, \v Z., Elitzur M., 2002, ApJ 570, L9

P\'erez Garc\'{\i}a A.M., Rodr\'{\i}guez Espinosa J.M., Santolaya Rey A.E.,
 1998, ApJ 500, 685

Rowan-Robinson M., 1995, MNRAS 272, 737

Scoville N.Z., Frayer D.T., Schinnerer E., et al., 2003, ApJ 585, L105

Soifer B.T., Houck J.R., Neugebauer G., 1987, ARAA 25, 187
 
Spinoglio L., Andreani P., Malkan M.A., 2002, ApJ 572, 105

Van Bemmel I.M., Barthel P.D., de Graauw M.W.M., 2000, A\&A 359, 523

Van Bemmel I.M., Barthel P.D., 2000, A\&A 379, L21

Van Bemmel I.M., Dullemond C.P., 2003, A\&A 404,~1

Walter F., Bertoldi F., Carill C., et al., 2003, Nature 424, 406

Yun M.S., Carilli C.L., 2002, ApJ 568, 88

\end{document}